# Prediction of Fund Net Value Based on ARIMA-LSTM Hybrid Model

Peng Zhou
Chengdu University of Technology
Chengdu, China
zhou.peng@student.zy.cdut.edu.cn

Fangyi Li
*Chengdu University of Technology*
Chengdu, China
li.fangyi@student.zy.cdut.edu.cn

*Abstract*—The net value of the fund is affected by performance and market, and the researchers try to quantify these effects to predict the future net value by establishing different models. The current prediction models usually can only reflect the linear variation law, poorly handled or selectively ignore their nonlinear characteristics, so the prediction results are usually less accurate. This paper uses a fund prediction method based on the ARIMA-LSTM hybrid model. After preprocessing the historical data, the first filter out the linear data characteristics with the ARIMA model, then pass the data to the LSTM model to extract the nonlinear characteristic by residual, and finally superposition the respective prediction values of the two models to obtain the prediction results of the hybrid model. Empirically shows that the methods in the paper are more accurate and applicable than traditional fund prediction methods.

## I. INTRODUCTION

With investment and financial education into the national education system, public awareness of financial management gradually increases [1]. At the same time, due to the epidemic's impact, the market interest rate continues to decline, making the fund market with low-risk high returns one of the most favoured financial options for investors. Data show that the average yield of 2020 partial stock hybrid funds was nearly 58%, and the average equity funds reached 60%, both of which are the highest level in nearly 11 years. The fund is quite active in market trading with its excellent performance. For the fund investment, the change of the fund's net value is an important factor in measuring the fund's profitability. Predicting the fund's income through the net value has also become a hot research topic in recent years. In this context, some experts and scholars have launched a wealth of research on this problem. Chen Jianing predicted the income of Internet money fund through wavelet analysis [3], Xiang Ying and others used the ARIMA model to predict the fund net value [4], and Meng Guoying tried to apply machine learning to fund performance prediction [5]. However, because various factors influence the fund in the market, most of these models only pursue the linear law of the net value change, ignoring the impact of the nonlinear change of the fund's net value on the prediction accuracy. Based on this, because the ARIMA model can handle the linear features in the time series well, and the LSTM model has an excellent performance in dealing with nonlinear problems [6], this paper proposes a hybrid model called ARIMA-LSTM that combines the two models. After analyzing the fund's net value trend, the prediction model obtained is compared with the traditional single model. The experimental results found that the ARIMA-LSTM hybrid model has better fitting performance and higher prediction accuracy than the traditional single model in applying fund net value prediction.

## II. ALGORITHM PRINCIPLE

### 2.1 ARIMA model

Autoregressive Integrated Moving Averaged Model is a commonly used time series forecasting method. The core idea of the model is to find a suitable mathematical function to fit the linear relationship between the current time value, the past time value, and the random interference amount to infer the future value through the past value [7]. The essence of the ARIMA model is an improvement of the ARMA model, and its mathematical formula is:

$$A_t = \varphi_1 A_{t-1} + \varphi_2 A_{t-2} + ... + \varphi_p A_{t-p} + \varepsilon_t - \theta_1 \varepsilon_{t-1} - \theta_2 \varepsilon_{t-2} - ... - \theta_z \varepsilon_{t-z}$$
$$+ \varepsilon_t - \theta_1 \varepsilon_{t-1} - \theta_2 \varepsilon_{t-2} - ... - \theta_z \varepsilon_{t-z} \quad (1)$$

Where $\varepsilon_t$ is the residual and $A_t$ is a stable time series. It should be noted that the ARIMA model can process only stable time series. If the time series is unstable, it needs to be transformed into a stable series by difference. The processed model is called the ARIMA model, denoted as $ARIMA(p,d,q)$.

Both p and q represent the order of the model. When p=0, the model degenerates to a q-order MA model MA(q), and when q=0, the model degenerates to a p-order AR model AR(p). In addition, d indicates how many differences have passed.

The basic steps of ARIMA model modelling are: analyzing the ADF value of the series, determining the (p, d, q) value of the model, estimating the correlation coefficient of MA and AR, testing the white noise series, and creating a prediction model.

*2.2 LSTM model*

In order to solve the exploding gradient and gradient disappearance of the recurrent neural network (RNN) during the operation, Hochreiter and Schmidhuber proposed an improved method for the recurrent neural network, namely the LSTM neural network model (Long Short-Term Memory) [8]. Unlike the RNN model, the LSTM model resets a cell state in the original hidden layer to preserve long-term memory. The LSTM structure is shown in Figure 1. The internal structure of the model is mainly composed of three control gates: input gate, forget gate and output gate. It is worth noting that tanh is the activation function, $C_{t-1}$ and $C_t$ represent the cell state at $t-1$ and $t$, respectively. $h_t$ and $h_{t-1}$ are the hidden states of the cell at $t$ and $t-1$.

FIGURE 1: STRUCTURE DIAGRAM OF LSTM MODEL

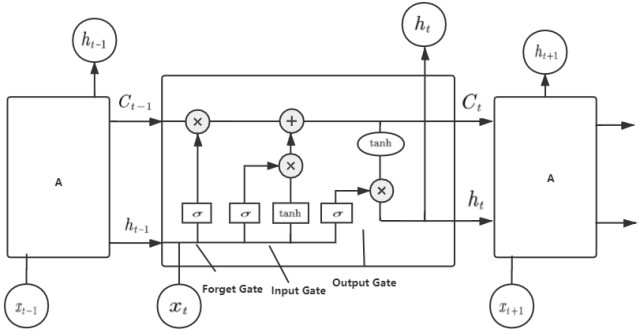

Firstly, the hidden state $h_{t-1}$ at the time can be determined through the forget gate of the model, and the degree of information retention of input $x_t$ can also be determined. The formula is:

$$f_t = \sigma(W_f \cdot [h_{t-1}, x_t] + b_f) \quad (2)$$

Secondly, you can determine how much content in the input variable can be stored in the cell state $C_t$ through the input gate. The formula is:

$$i_t = \sigma(W_i \cdot [h_{t-1}, x_t] + b_i) \quad (3)$$

$$\widetilde{C}_t = \tanh(W_c \cdot [h_{t-1}, x_t] + b_c) \quad (4)$$

$$C_t = f_t \times C_{t-1} + i_t \times \widetilde{C}_t \quad (5)$$

Finally, the output gate of LSTM outputs the hidden state of each cell, the formula is:

$$O_t = \sigma(W_o \cdot [h_{t-1}, x_t] + b_o) \quad (6)$$

$$h_t = O_t \cdot \tanh(C_t) \quad (7)$$

In the above formula, $W_f$, $W_i$, $W_c$, $W_o$ are the weight matrix of different control gates; $b_f$, $b_i$, $b_c$, $b_o$ are the bias term of each control gate; $C_t$ and tanh are the corresponding activation functions, which express how much information passes through the different control gates.

## III. ARIMA-LSTM FORECASTING MODEL

Changes in fund net worth usually have strong nonlinear and irregular [9], and predictions using only a single model often yield poor results. Based on the fund's net value, using the ARIMA-LSTM hybrid model, filter the linear features with the ARIMA model, and then give the nonlinear characteristics stored to the LSTM model for processing, which can ensure the linear and nonlinear characteristics of the data. Finally, combining the prediction results of two models to obtain the prediction results of the hybrid model. See Figure 2 for its flowchart.

FIGURE 2: PREDICTION FLOW CHART OF ARIMA-LSTM HYBRID MODEL

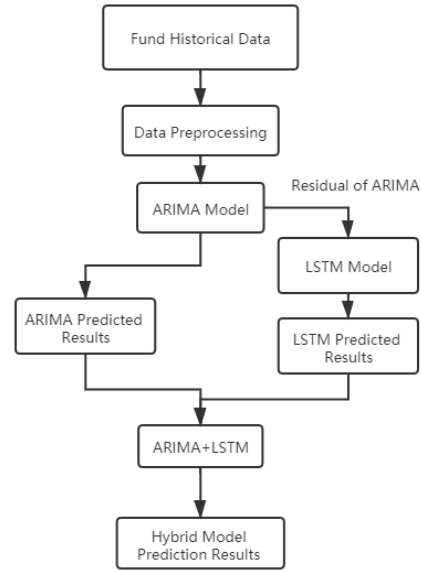

The time series $y_t$ of the fund's net value can be regarded as consisting of a linear structure $L_t$ and a non-linear structure $N_t$. The mathematical formula is:

$$y_t = L_t + N_t \quad (8)$$

First, we use the ARIMA model to predict the linear part of the series and get the predicted value $\widehat{L}_t$. Then subtract $\widehat{L}_t$ from the true value to get the residual series $e_t$.

$$e_t = y_t - \widehat{L}_t \quad (9)$$

Use the LSTM model to process the series obtained in the previous step to predict the non-linear part of the fund's net value to obtain the predicted value $\widehat{N_t}$.

$$\widehat{N_t} = f(e_{t-1}, e_{t-2}, ..., e_{t-m}) + \varepsilon_t \quad (10)$$

Finally, the ARIMA-LSTM hybrid model's predicted value equals the sum of the two-step predicted values.

$$\widehat{y} = \widehat{L_t} + \widehat{N_t} \quad (11)$$

## IV. EXAMPLE ANALYSIS

### 4.1 data processing

This article selects the 1260-day fund net value data of Huabao Hybrid Fund (240008) from June 6, 2016, to July 30, 2021. The data used is derived from the historical net value of funds that have been published on the Tiantian Fund Website. The 1260-day data is divided into three parts, as shown in Table 1, for different model training processes.

TABLE 1: CLASSIFICATION OF TRAINING DATA

| Code | Size | Train Size | Val Size | Test Size |
|---|---|---|---|---|
| 240008 | 1260 | 900 | 100 | 260 |

At the same time, this paper adopts the sliding window prediction method [10], as shown in Figure 3. Assuming L is the length of the window to be trained, starting from the leftmost time t, this model predicts the next days' net value and continues to move forward one day until it reaches the rightmost time T. It should be noted that the model will only predict the net value on the next day and will use L-length historical data for analysis before predicting.

FIGURE 3: SLIDING WINDOW ONLINE PREDICTION METHOD

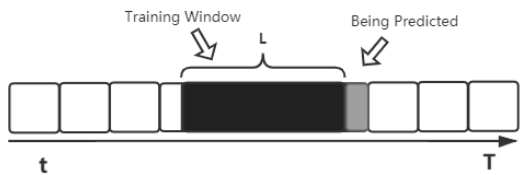

### 4.2 Evaluation indicators

This article selects three common error evaluation indicators to evaluate the prediction accuracy of different models [11]. These three indicators are MSE, MAE, RMSE, and the following are their respective mathematical expressions:

$$MSE = \frac{1}{n} \sum_{i=1}^{n} \left(\widehat{y_i} - y_i\right)^2 \quad (12)$$

$$MAE = \frac{1}{n} \sum_{i=1}^{n} \left|\widehat{y_i} - y_i\right| \quad (13)$$

$$RMSE = \sqrt{\frac{1}{n} \sum_{i=1}^{n} \left(\widehat{y_i} - y_i\right)^2} \quad (14)$$

The three indicators have the following characteristics: the smaller the value, the smaller the error between the value predicted by the model and the true value, which means the higher accuracy.

### 4.3 Result analysis

#### 4.3.1 Constructing ARIMA model

Figure 4 is the original time series chart of the fund. It is not difficult to see that the data changes drastically, and there is no obvious change rule. Besides, the chart rises sharply after 1000 days, proving that the series is a non-stable series, so the difference method is needed to convert the original sequence into a stable series. Figure 5 is the series diagram after the first-order difference processing. It can be seen that the processed series is more stable than the original series. Through the ADF test, it can be determined that the series can become stable after the first-order difference (d=1)

FIGURE 4: TIME SERIES DIAGRAM OF FUND NET VALUE

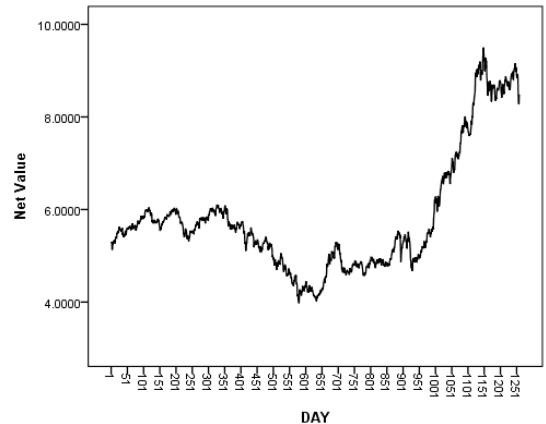

FIGURE 5: TIME SERIES DIAGRAM AFTER FIRST-ORDER DIFFERENCE

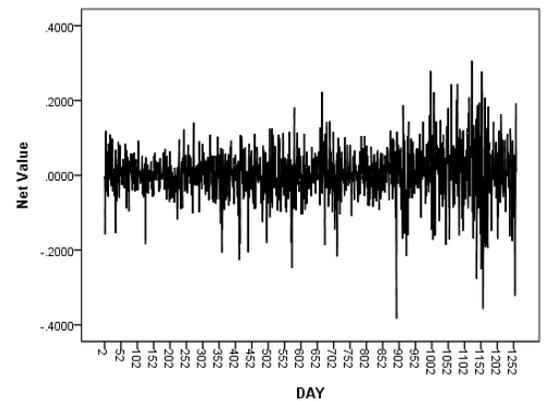

Then the parameters p and q of the ARIMA model can be inferred from the autocorrelation coefficient (ACF) and partial autocorrelation coefficient (PACF) diagrams of the series. Figures 6 and 7 are the ACF and PACF diagrams of the fund's net value. After analysis, it is found that the ACF diagram is lagging the first-order truncation, and the PACF diagram is also the first-order truncation. In order to improve the accuracy of the model, this article refers to the AIC values of different (p, d, q) combinations, and the AIC value of the combination (0, 1, 0) is the smallest, which is -2851, indicating the model fit created under this combination Highest degree and best prediction effect.

FIGURE 6: ACF DIAGRAM

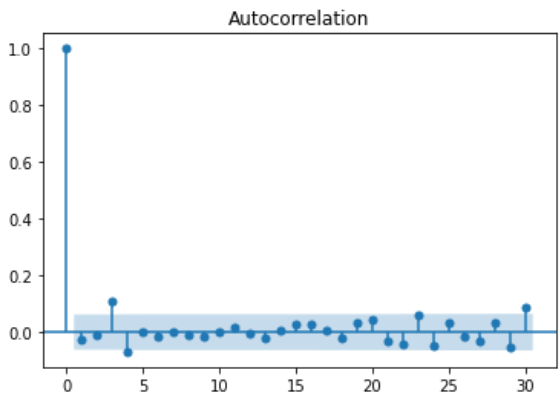

FIGURE 7: PACF DIAGRAM

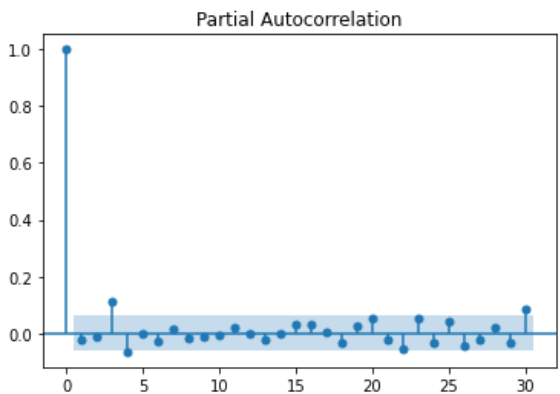

According to the experimental data, the prediction formula of the ARIMA model (p=0, d=1, q=0) can be written:

$$(1-L)y_t = \alpha_0 + \varepsilon_t \qquad (15)$$

$$y_t = \alpha_0 + y_{t-1} + \varepsilon_t \qquad (16)$$

If only use ARIMA single model to predict the fund's net value, the obtained model prediction chart is shown in Figure 8. It can be seen that the ARIMA model has a low prediction accuracy of the fund's net value after 1000 days, which means the model is not suitable for use in actual fund forecasting

FIGURE 8: NET VALUE FORECAST CHART OF ARIMA MODEL

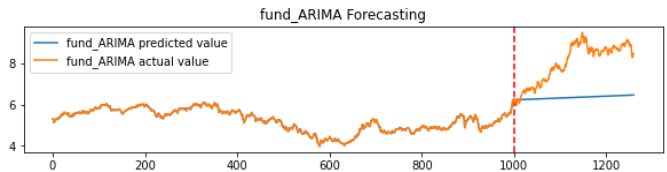

### 4.3.2 LSTM processing ARIMA residuals

Finally, use Python to build a suitable LSTM model. After many parameter adjustments, the final model structure is determined as follows: the number of layers is 3, the input and output dimensions are both set to 1, the learning rate is 0.005, the training iteration is 100 times, and the batch_size is 64. Figure 9 and Figure 10 show the fit of the target fund and the forecast of future net value using the hybrid model. It can be seen that the value predicted by the ARIMA-LSTM hybrid model is roughly the same as the real trend, and the degree of fit is significantly better than that of the ARIMA model.

FIGURE 9: FITTING DIAGRAM OF ARIMA-LSTM HYBRID MODEL

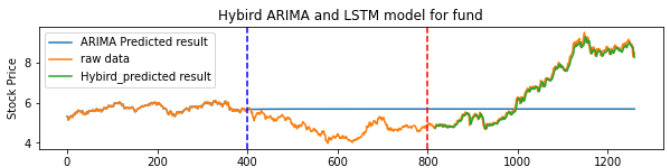

FIGURE 10: PREDICTION DIAGRAM OF ARIMA-LSTM HYBRID MODEL

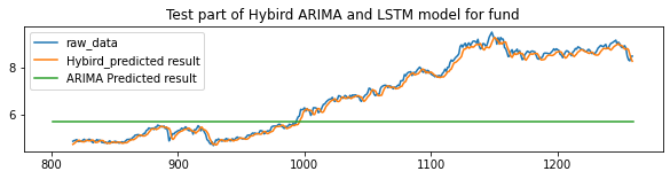

### 4.3.3 Comparison of prediction indicators of three models

Under the premise of the same research data, the prediction results of the three models are shown in Table 2. It can be seen from the data that the values of MSE, MAE, and RMSE of the LSTM model are much lower than those of the ARIMA model, and the values of the ARIMA-LSTM hybrid model are also lower than those of the other two single models. Through the quantitative analysis of the indicators, the paper found out that the prediction effect of the ARIMA model is the worst, the prediction effect of the LSTM model is better, the prediction effect of the hybrid model is the best. To sum up: the ranking of the prediction effects of the three models from high to low is the ARIMA-LSTM hybrid model, the LSTM model and the ARIMA model. In conclusion, the ARIMA-LSTM hybrid model is a more reliable time series analysis model, which is more suitable for predicting the fund's net value in real life than the independent model.

TABLE 2: ERROR INDICATORS TABLE OF PREDICTION RESULTS

| Model | MSE | MAE | RMSE |
|---|---|---|---|
| ARIMA | 3.61 | 1.69 | 1.90 |
| LSTM | 0.13 | 0.31 | 0.36 |
| ARIMA-LSTM | 0.01 | 0.09 | 0.12 |

## V. CONCLUSION

The change of the fund's net value has both linear and non-linear characteristics. Using traditional models to predict the net value is difficult to deal with non-linearity, resulting in low accuracy of prediction. Although machine learning prediction methods have great advantages in dealing with non-linear problems, they are prone to overfitting when dealing with small data samples, making the prediction accuracy not high. The hybrid model separates the two characteristics and combines their respective advantages. It performs well in dealing with complex time series issues such as the fund's net value and has proved to be a more reliable analysis and forecast tool.